# Spin-Phonon Coupling and High Pressure Phase Transitions of RMnO$_3$ (R= Ca and Pr): An Inelastic Neutron Scattering and First Principle Studies

S. K. Mishra[1], M. K. Gupta[1], R. Mittal[1], A. I. Kolesnikov[2] and S. L. Chaplot[1]

[1]*Solid State Physics Division, Bhabha Atomic Research Centre, Mumbai 400085, India*
[2]*Chemical and Engineering Materials Division, Oak Ridge National Laboratory, Oak Ridge, Tennessee 37831, USA*

We report inelastic neutron scattering measurements over 7-1251 K in CaMnO$_3$ covering various phase transitions, and over 6-150 K in PrMnO$_3$ covering the magnetic transition. The excitations around 20 meV in CaMnO$_3$ and at 17 meV in PrMnO$_3$ are found to be associated with magnetic origin. In spite of similarity of the structure of the two compounds, the neutron inelastic spectrum of PrMnO$_3$ exhibits broad features at 150 K unlike well-defined peaks in the spectrum of CaMnO$_3$. This might result from the difference in nature of interactions in the two compounds (magnetic and Jahn-Teller distortion). The interpretation and analysis of the observed phonon spectra have been performed using ab-initio phonon calculations. We also discuss the effect of pressure on the structural distortions in the orthorhombic phase of CaMnO$_3$ and PrMnO$_3$. On application of pressure, we found that the variations of Mn-O distances are isotropic for CaMnO$_3$ and highly anisotropic for PrMnO$_3$. The calculated structure as a function of pressure in PrMnO$_3$ shows that suppression of Jahn-Teller distortion and insulator to metal transition occurs simultaneously. Our calculations show that this transition may not be associated with the occurrence of the tetragonal phase above 20 GPa as reported in the literature, since the tetragonal phase is found to be dynamically unstable although it is found to be energetically favored over the orthorhombic phase above 20 GPa. CaMnO$_3$ does not show any phase transition up to 60 GPa.

**PACS numbers:** 78.70.Nx, 63.20.-e, 61.50.Ks, 65.40.-b

**Keywords**: Neutron inelastic scattering, Phonons in crystal lattice, Crystallographic aspects of phase transformations; pressure effects, Thermal properties of crystalline solids

I. INTRODUCTION

The perovskite transition-metal oxides show a variety of interesting physical properties, such as dielectric, magnetic, optical, and transport properties [1-10]. The study of perovskite manganite $RMnO_3$ (R= Ca, La, Pr etc) has been of great relevance [6-17]. Among these, $CaMnO_3$ has received much interest in the past decade due to its structural [18], physical [19], magnetic [20], and thermo-electrical properties [21,22].

$CaMnO_3$ crystallizes in the distorted orthorhombic structure (space group Pnma) and consists of single-valent $Mn^{4+}$ ions and does not exhibit Jahn-Teller distortion. At ambient conditions, it is paramagnetic and at ~130 K it undergoes a G type antiferromagnetic (AFM) transition [23]. In high temperature x-ray diffraction study, H Taguchi *et al* [24] showed that oxygen deficient $CaMnO_3$ undergoes orthorhombic to tetragonal phase at 1169 K and finally transforms in cubic phase at 1186 K. Doping of trivalent and tetravalent ions at $Ca^{2+}$ site provides additional electrons into $e_g$ orbital of Mn ions and causes a variety of phase transitions, e.g from collinear G-type AFM insulator to canted G-type AFM metal, collinear C-type AFM insulator etc. This suggests that we can tune the magnetic phase transition with a small amount of electron doping. For example, solid solution of $(1-x)CaMnO_3$-$xPrMnO_3$ ($Ca_{1-x}Pr_xMnO_3$) exhibits a variety of ground state, namely: G- AFM insulator, CE- type charge ordered state, CE-type AFM spins ordering, canted AFM state which consists a mixture of AFM and ferromagnetic (FM) clusters, depending on temperature and composition [25-28]. The other end member of solid solution, $PrMnO_3$ also crystallizes in the orthorhombic phase but it exhibits [16] a strong Jahn-Teller distortion of $MnO_6$ octahedral associated with the ordering of $e_g$ orbital and undergoes A-type antiferromagnetic spin ordering below 95 K [16]. The difference in magnetic structure of $CaMnO_3$ and $PrMnO_3$ is attributed to different occupancy of d electron in Mn ions. Hence it is intriguing to study the role of magnetic interaction on lattice vibration.

The structural, physical and magnetic properties of $RMnO_3$ (R= Ca and Pr) were investigated by variety of experimental techniques [29-36]. The physical properties of these materials are governed by a delicate interplay of charge, spin, orbital, and lattice degrees of freedom. To understand the relation among these interactions, first principles calculations have been performed in cubic and orthorhombic phases. Using a density functional theory approach within the local spin density approximation, S. Bhattacharjee *et al*, [33] have computed structural, dielectric and dynamical properties of orthorhombic phase of $CaMnO_3$. They have computed whole set of zone-center phonon modes and assignment of experimental data has been proposed. F.P. Zhang *et al* [34] had investigated geometry, ground state electronic structure and charge distributions of CaMnO$_3$. Effect of compressive and tensile strains on a magnetic phase transition in electron-doped CaMnO$_3$ was theoretically studied by H. Tsukahara *et al* [35]. Further the first principles calculations of structural, electronic and magnetic properties of the PrMnO$_3$ as well as the calculation of zone centre phonon modes in the cubic phase were reported by B. Bouadjemi [36].

Understanding of various functional properties of derived compounds Pr$_x$Ca$_{1-x}$MnO$_3$ needs complete study of electronic and dynamical properties of the parent compound CaMnO$_3$ and PrMnO$_3$. In spite of various experimental and theoretical investigations, the temperature dependence of phonon spectra for CaMnO$_3$ and PrMnO$_3$ is still missing. In the present study, we report results of inelastic neutron scattering measurements of phonon spectra at different temperatures of CaMnO$_3$ and PrMnO$_3$. The interpretation and analysis of the observed phonon spectra have been performed using ab-initio phonon calculations. Previous first-principles calculations on CaMnO$_3$ were focused mainly on the electronic structure [33-36]. Here, we report a first-principles study of phonon dynamical properties of the ground-state orthorhombic phase in the entire Brillouin zone. We also investigated the effect of pressure on the structural distortions in orthorhombic phase of CaMnO$_3$ and PrMnO$_3$. The calculated structure as a function of pressure in PrMnO$_3$ shows that suppression of Jahn-Teller distortion and insulator to metal transition occurs simultaneously. We provide the evidence that although tetragonal phase as reported in the literature is energetically favored over orthorhombic phase above 20 GPa, but it is found to be dynamically unstable.

## II. EXPERIMENT

Polycrystalline samples of CaMnO$_3$, PrMnO$_3$ were prepared by a solid state reaction method using powders of CaCO$_3$, Pr$_6$O$_{11}$ and MnO$_2$ each of minimum 99.5% purity. These powders were thoroughly mixed in stoichiometric amounts using zirconia jar and balls in a ball mill using acetone as the mixing medium. Calcination of the mixed powder was carried out at different temperature and times with several intermediate grindings. The calcined powder was characterized using x-ray diffraction and temperature and times was optimized for single phase purity. Single phase calcined powder was pressed into circular pellets of 12 mm diameter and ~1.25 mm thickness using a uniaxial hydraulic press at an optimized load of 100 kN. Sintering of the green pellets was carried out at 1700 K for 20 hours in air. For powder x-ray diffraction experiments, the sintered pellets were crushed to fine powder and subsequently annealed at 750 K to remove strains introduced, if any, during crushing. The phase purity of sample was confirmed by x-ray powder diffraction recorded using Cu Ka radiation in the angular range 20 -130 degrees at room temperature.

Inelastic neutron scattering measurements were performed using the Fine Resolution Chopper Spectrometer (SEQUOIA) [37,38] at the Spallation Neutron Source (SNS) at Oak Ridge National Laboratory. The data were collected using fixed incident neutron energy of 110 meV, which allowed for the measurement of excitations up to energy transfers of ~100 meV. The low temperatures scans (4-300 K) were carried out using the closed cycle helium refrigerator. For high temperature measurements up to 1250 K samples were heated in air. The measurements were done with high energy resolution for low temperature (T<305 K) measurements of CaMnO3, providing energy resolution at the elastic line $\Delta E$=2.0 meV. All other measurements were done in high intensity configuration, with energy resolution at elastic line $\Delta E$=5.5 meV. The resolution in both settings becomes better with increasing energy transfer and around 80 meV it reaches about 1/3 of that at the elastic line. We have used furnace (called MiCAS) with a quartz tube insert that allows controlling the atmosphere of sample.

In the incoherent one-phonon approximation the scattering function $S(Q,E)$, as observed in the neutron experiments, is related to the phonon density of states [39] as follows:

$$g^{(n)}(E) = A < \frac{e^{2W(Q)}}{Q^2} \frac{E}{n(E,T) + \frac{1}{2} \pm \frac{1}{2}} S(Q,E) > \quad (1)$$

$$g^n(E) = B \sum_k \{\frac{4\pi b_k^2}{m_k}\} g_k(E) \quad (2)$$

where $Q$ and $E$ are neutron momentum and energy transfers, respectively, the + or − signs correspond to energy loss or gain of the neutrons respectively and $n(E,T) = \exp(E/k_B T) - 1^{-1}$. $A$ and $B$ are normalization constants and $b_k$, $m_k$, and $g_k(E)$ are, respectively, the neutron scattering length, mass, and partial density of phonon states of the $k^{th}$ atom in the unit cell. The quantity within < ---- > represents suitable average over all $Q$ values at a given energy. $2W(Q)$ is the Debye-Waller factor averaged over all the atoms. The weighting factors $\frac{4\pi b_k^2}{m_k}$ for various atoms in the units of barns/amu are: Ca: 0.071; Pr: 0.019; Mn: 0.039; O: 0.264. The MANTIDPLOT [40] and DAVE [41] software packages were used for data reduction and analysis. The data were collected over a wide range of Q, 1-7 Å$^{-1}$. Since both the compounds contain magnetic atoms, contributions from magnetic scattering are also present in the measured data.

## III. THEORETICAL CALCULATIONS

The calculation of phonon spectra in the *Pnma* space group has been performed using the first-principles density functional theory and both the local density approximation (LDA) or generalized gradient approximation (GGA) for $CaMnO_3$. However, for $PrMnO_3$, GGA has been used. Relaxed geometries and total energies were obtained using the projector-augmented wave formalism [42-43] of the Kohn-Sham formulation of the density-functional theory [44-45] within both the LDA and GGA level, implemented in the Vienna *ab initio* simulation package (VASP) [46]. The GGA was formulated by the Perdew-Burke-Ernzerhof (PBE) density functional [47] and LDA was based on the Ceperly–Alder parametrization by Perdew and Zunger [48]. All results are well converged with respect to *k* mesh and energy cutoff for the plane-wave expansion. The total energy calculations have been done using an energy cutoff of 1100 eV. A grid of 8×8×8 K-points was used according to the Monkhorst-Pack (MP) scheme [49]. The convergence criteria for the self-consistent field and for the ionic relaxation loops were set to $10^{-8}$ eV Å$^{-1}$ and $10^{-4}$ eV Å$^{-1}$, respectively. Hellmann-Feynman forces following geometry optimization were less than $10^{-4}$ eV Å$^{-1}$. We have used the G-type and A-type antiferromagnetic structures in orthorhombic phases for $CaMnO_3$ and $PrMnO_3$ respectively. The choice of magnetic structures is based on the previously reported configurations obtained from neutron diffraction measurements [16, 23]. However in the tetragonal phase of $PrMnO_3$ the magnetic structure has not been known. So we have calculated the enthalpy of the tetragonal phase of $PrMnO_3$ including different antiferromagnetic configurations, namely A, C, G type, and ferromagnetic configurations. The calculations show that ferromagnetic structure is favored in comparison to the other structures. In the cubic phase G-type antiferromagnetic and ferromagnetic structures are adapted for $CaMnO_3$ and $PrMnO_3$ respectively.

The supercell approach or direct method is used to calculate the phonon frequencies in entire Brillouin zone implemented in PHONON5.12 software [50]. The supercell of dimension 2×2×2 has been used to calculate the Hellman-Feynman forces in both the compounds in orthorhombic phase and tetragonal phase for $PrMnO_3$. Phonon calculation for high temperature cubic phase Pm-3m, which has a small unit cell containing 5 atoms, we have chosen a supercell of dimension 4×4×4 to properly account the long range interaction in both the compounds. The forces were extracted from individual displacements of the symmetry inequivalent atoms in the super cell along three Cartesian directions ((±x, ±y, ±z). In orthorhombic phases 24 distinct configurations are generated by displacing the symmetrically inequivalent atoms in both the compounds. However in tetragonal and cubic phase 16 and 8 respectively distinct displacements are found.

The phonon calculations are performed in various magnetic and nonmagnetic configurations to see the effect of spin degrees of freedom on phonons. These calculations enable us to quantify the strength of the coupling between spin and phonons if any. The nonmagnetic calculations are done in fully relaxed and partially relaxed geometry. In the fully relaxed calculations the atomic coordinates and lattice parameter have been relaxed. However partially relaxed nonmagnetic calculation (PNM) refers to fix lattice parameter as used in magnetic calculations with relaxing the atomic coordinates. Hereafter, the labeling "FRM" and "FRNM" refer to fully relaxed magnetic and fully relaxed nonmagnetic calculation.

## IV. RESULTS AND DISCUSSION

**(a) Temperature dependence of neutron inelastic spectra in orthorhombic phase of $CaMnO_3$**

$CaMnO_3$ is known to undergo paramagnetic to antiferromagnetic (AFM) transition at ~ 130 K [23]. Other transitions are observed from orthorhombic to tetragonal phase at 1169 K and then to cubic phase at 1186 K. The phonon spectra of $CaMnO_3$ (Figs. 1-3) were measured from 7 up to 1250 K, across the magnetic transition (~130 K) and the structural phase transitions at high temperatures. The neutron inelastic spectra were collected over wide range of momentum transfer (Q) from 0.5-7 Å$^{-1}$. The magnetic signal is expected to be more pronounced at low Q, and it decreases as Q increases, following the magnetic form factor. Hence we have integrated the inelastic scattering function S(Q,E) in two ranges, namely ~~low~~ total Q range (0.5-7 Å$^{-1}$) and high Q (4-7 Å$^{-1}$), the later one represents mostly the contribution from the neutron scattering on phonons.

The data collected at 7 K show intense peak at about 20 meV in the low Q data (Figs. 1, 2). The observed peak around 20 meV is largely contributed from magnetic excitations and shows strong Q dependence intensity (stronger at low-Q and weaker at high-Q). However as the temperature increases the contribution from spin excitation becomes weaker. Above the magnetic transition temperature ($T_N$=130 K) we also see that the intensity of peak at 20 meV is decreasing with increase in temperature up to 601 K (Fig. 3). At temperatures 951 K and above the intensity does not change with increase in temperature. This indicates that the contributions from the paramagnetic scattering persist up to very high temperatures of 601 K.

Fig. 2 illustrates the experimental S(Q,E) measured for CaMnO$_3$ in the low temperature range from 7 to 300 K. The data collected at 7 K clearly show the signature of spin-wave excitations which are probably gaped at Q=1.45 Å$^{-1}$, corresponding to Bragg peak (011) (the gap is not resolved with the current experimental setup, providing energy resolution ~2 meV at the elastic line). Further we noticed that at higher temperature (T=110 K) the intensity of these excitations are strongly redistributed to lower energies, the peak position shifts from 19.5 meV at 7 K to 16.5 meV at 110 K. At T>T$_N$ (in paramagnetic state) at small Q values (Q<3.5 Å$^{-1}$) we observe spin-fluctuation excitations extended from elastic line up to ~20 meV. It is evident from figure 2 that the nature of neutron scattering intensity S(Q, E) changes above T$_N$ from inelastic (due to well defined spin excitation) to quasielastic (due to stochastic spin- fluctuations) and the intensity also spreads over a broader range of Q. The intensity of these spin fluctuations corrected for the thermal population Bose factor leads to a peak in g(E), which is plotted in Figs. 1 & 3, that decreases with temperature increase.

The high temperature neutron inelastic spectra (Fig. 3) show strong modification with increasing temperature. In particular, the intensity of peaks below 40 meV decreases with increasing temperature up to 601 K while the peak around 90 meV shows significant shift towards lower energies with increasing temperature. The others peaks do not change in a noticeable way. We find that above 1101 K, a prominent change is observed in the neutron inelastic spectra, where they are found to be more diffusive as compared to the spectra at low temperatures. These changes in the phonon spectra may be associated with phase transitions in CaMnO$_3$ which transforms from orthorhombic to tetragonal phase at 1169 K, and finally to cubic phase at 1186 K [24].

**(b) Magnetic ordering and calculated phonon spectra in the orthorhombic phase of CaMnO$_3$**

At room temperature AMnO$_3$ crystallizes in the orthorhombic phase (*Pnma*) with 4 formulas per unit cell (20 atoms). The irreducible representations of the zone-centre optical phonons are

$$\Gamma_{optical} = (7A_g + 5B_{1g} + 7B_{2g} + 5B_{3g})_{Raman-active} + (9B_{1u} + 7B_{2u} + 9B_{3u})_{IR-active} + (8A_u)_{silent}.$$

Figure 4 compares the computed zone centre phonon frequencies with experimental and theoretical data available in the literature for both the compounds. Our computed data are found to be in fair agreement with reported data, which validates the computations.

In order to analyze the experimental data, as stated above, first principle calculations have been performed. The fully optimized structure of CaMnO$_3$ in both the "FRM" and "FRNM" configurations using both the local density approximation (LDA) or the generalized gradient approximation (GGA) exchange correlation functional are summarized in Table I. The calculated atomic positions are in good agreement with the experimental data. It can be seen that in both the LDA and GGA, the "FRM" calculated structures are found to close to the experimental data [51]. The LDA calculated lattice constants are found to be underestimated by 2% in comparison to the experimental data, while GGA gives slight overestimation of about 1%. As will be shown latter in the paper, nonmagnetic structures are found to be dynamically unstable at T=0 K.

The comparison between the experimental data and the calculated phonon spectra from both the LDA and GGA are shown in Fig. 5. It can be seen that all the observed features in the experimental data are fairly well reproduced by the calculations. It is evident from this figure that below 55 meV, phonon calculation in fully relaxed magnetic configuration using LDA gives better agreement with the experimental data in comparison to GGA. However, for the spectral range above 55 meV, the calculated phonon spectrum using GGA describes the experiment better than LDA. The high energy (> 75 meV) phonons are due to Mn-O stretching modes. The observed discrepancy could be understood in terms of bond lengths. As shown in TABLE I, the LDA calculation underestimates the lattice parameters. The shorter bond lengths would shift the phonon spectra to higher energies in comparison to the experimental data. Similarly the slight overestimation of lattice parameter by GGA, results in underestimation of the energies of the Mn-O modes in the calculation.

The atomistic contributions in the phonon spectra from the various calculations can be understood in terms of the partial density of states. The difference is primarily due to the nature of the chemical bonding in the magnetic and nonmagnetic configurations, as well as the related volume effect. The computed atomistic partial phonon density of states show (Fig. 6 (a)) that the contribution of the oxygen atoms spreads over the whole energy range, while the Mn atoms contribute mainly up to 75 meV. The contribution due to Ca atoms extends up to 50 meV. Above 75 meV, the dynamics is mainly due to the Mn-O stretching modes. It can be seen (Fig. 6 (a)) that the contributions of Ca and Mn are nearly the same in the calculations performed using LDA and GGA, while there are significant differences in the partial contributions from the O atoms. The difference in the calculated partial contribution is mainly in the stretching modes region which is very sensitive to the unit cell volume. The underestimation or overestimation of energy of modes is related to the calculated structures as given in TABLE I. Results of the FRM-GGA calculations are found to be close to the experimental data. Consequently in the following we adopt the GGA density functional.

To study the effect of magnetic interactions on the phonon spectra, the phonon spectra were calculated in three different configurations as said above ("FRM", "FRNM", and "PNM") and shown in Figure 6 (b). It is evident from this figure that the phonon spectra calculated using PNM results in several unstable modes up to $25i$ meV. Similarly FRNM calculations also results in unstable modes with slightly lower energy range ($10i$ meV). The comparison of the FRM and PNM calculations indicates that magnetic interactions are very important for obtaining the dynamically stable structure.

**(c) Temperature dependent neutron inelastic spectra and First principle studies in orthorhombic phase of $PrMnO_3$**

The compound $PrMnO_3$ is isostructural to $CaMnO_3$; however, the magnetic structure is different. The Jahn-Teller transition in $PrMnO_3$ takes place at about 1050 K. It undergoes antiferromagnetic spin ordering below 95 K. Here we performed the measurements of neutron inelastic spectra of $PrMnO_3$ at two temperatures, 6 and 150 K. The experimental S(Q,E) data for $PrMnO_3$ (Fig. 7, at the bottom) collected at 6 K clearly shows the signature of spin-wave excitations at energies below 20 meV and Q<1.5 Å$^{-1}$, which significantly dissipate at 150 K. This behavior is similar to $CaMnO_3$, however the Q dependence is completely different. Fig. 7 (at the top) shows the phonon spectra, (a) summed over momentum transfer (Q)=1-7 Å$^{-1}$ and (b) Q=4-7 Å$^{-1}$. The data collected at 6 K show intense excitations with the peak maximum at about 17 meV. However above magnetic transition temperature the intensity at 17 meV peak is significantly suppressed but has contribution from phonons and paramagnetic fluctuations.

It is remarkable to notice that in spite of similar structure, room temperature neutron inelastic spectrum (Fig. 5) of $CaMnO_3$ exhibits well-defined peaks in comparison to the data for $PrMnO_3$ (Fig. 8) at 150 K. The difference in the phonon spectra may be due to the difference in nature of interactions. The Jahn-Teller distortion is present in $PrMnO_3$, while such distortion is not there in $CaMnO_3$.

The phonon calculations (Figs. 8 (a) and (b)) for $PrMnO_3$ are carried out only in the fully relaxed magnetic configuration using GGA exchange correlation functions. All the observed features are fairly well reproduced (Fig. 8(a)) by the computations. The computed atomistic partial density of states shows (Fig. 8(b)) that the contribution of the Mn and oxygen atoms spreads over the whole energy range up to 75 meV, while the contribution due to Pr atoms extends up to 50 meV. It should be noted that in the present case contributions due to the Mn-O stretching modes do not extend above 75 meV, while in CaMnO$_3$ the energy range of these modes is up to 90 meV. The comparison between the calculated and experimental phonon spectra shows that stretching modes are underestimated (Fig. 8) in the phonon calculations. Further the calculations of zone centre phonon modes as shown in Fig. 4 also confirm the same.

### (d) Structural instabilities in the cubic phase of CaMnO$_3$ and PrMnO$_3$

The cubic phase in both CaMnO$_3$ and PrMnO$_3$ occurs at very high temperatures. It is expected to be dynamically unstable at low temperature and stabilized by anharmonicity at high temperature. It is important to explain the unstable modes in order to understand the phase transition mechanism. In order to obtain the equilibrium lattice parameter for the cubic phase of CaMnO$_3$ and PrMnO$_3$, we performed structural optimization by minimizing the total energy with respect to structural parameter. The cubic phase of PrMnO$_3$ is known to be ferromagnetic and half metallic. The calculations for the cubic phase of CaMnO$_3$ and PrMnO$_3$ have been performed including the G- type antiferromagnetic and ferromagnetic ordering respectively. The relaxed lattice parameters for CaMnO$_3$ and PrMnO$_3$ are found to be a= 3.7588 Å and a= 3.8920 Å, respectively. Figure 9 show the computed phonon dispersion relation using generalized gradient approximation. It is evident from this figure that phonon dispersions in the two compounds show noticeable difference especially at higher energies. The phonons contribute up to 110 meV for the CaMnO$_3$ and up to 80 meV for the PrMnO$_3,$ respectively. This difference could be understood in terms of difference in bond lengths for these compounds. The lattice parameter of cubic CaMnO$_3$ is smaller than PrMnO$_3$ and hence shorter bond lengths would shift the phonon energies to higher side.

Careful inspection of phonon dispersion relation reveals the presence of polar instability (zone centre phonon instabilities) with an imaginary frequency $\omega_{FE} \approx 20i$ meV for CaMnO$_3$ and $\omega_{FE} \approx 8i$ meV in PrMnO$_3$ respectively. We also observed antiferrodistortive (AFD) instability (zone boundary phonon instabilities) at R and M points in the Brilloun zone for both compounds. Condensation of these AFD instabilities results in phase transition from cubic to the tetragonal and then to orthorhombic phase. It is also remarkable to notice that as we move from M to R point, the strength of instability is quite similar for CaMnO$_3$ ($\omega_{AFD} \approx 30i$ meV at R and M points), however, it becomes larger for PrMnO$_3$ ($\omega_{AFD} \approx 30i$ meV at R point and $\omega_{AFD} \approx 24i$ meV M point). Interestingly, the branches along the Γ-X, X-M, Γ-R, and Γ-M directions, show dramatic changes when reaching the X, R and M points. When moving away from R to M, two unstable modes are detected. One of them is rather flat and the other one shows rapid stiffening and becomes stable. Moreover, moving across the Γ-X, the unstable modes become stable at X point for $CaMnO_3$ and it does not show appreciable change for $PrMnO_3$. For $CaMnO_3$, Bhattacharjee et al [14] also observed both (polar and AFD) types of instability. They reported that the polar instability is very sensitive to volume and it is suppressed by anharmonic effects once the dominant antiferrodistortive (AFD) instability is condensed.

**(e) Cubic to tetragonal phase transition**

Now, we would like to comment on the phase transition mechanism of the cubic *(Pm-3m)* to tetragonal *(I4/mcm)* transition that occurs in $AMnO_3$ (A= Ca and Pr) at high temperature. Cubic to tetragonal phase transition is known [53] to be driven by R point phonon instability. To identify this specific phonon, we have calculated the phonon energies in the high temperature cubic phase of $AMnO_3$ with a super cell of √2×√2×2, which is equivalent to tetragonal phase (*I4/mcm*). The structure parameter used in phonon calculations is given in Table II. The distortion vector has been obtained with respect to ideal cubic phase. The calculated eigen vector of the unstable R point mode in the super cell (√2×√2×2) of cubic phase is also given in Table II. It is evident from the table that only O2 atoms displace from its ideal position. The eigen vector of the R point mode is in good agreement with the distortion vector. The calculations are in agreement with the previous results [14] that freezing of the unstable R point mode is responsible for cubic to tetragonal phase transition on lowering the temperature. The displacement pattern of unstable mode at the R point in the cubic phase is shown in Fig. 10, which indicates anti phase rotation of in plane oxygen's in the two layers containing $MnO_6$.

**(f) High pressure phase stability of $CaMnO_3$ and $PrMnO_3$**

Pressure tunes the interplay between lattice and electronic degree of freedom to a much larger extent than any other parameter like temperature and magnetic field. In the present section, we discuss the effect of pressure on the structural distortions in the orthorhombic phase of $CaMnO_3$ and $PrMnO_3$. The main emphasis of this study are to investigate the pressure dependence of the Jahn-Teller distortion (if persist) and existence of insulator to metal transition at high pressure. Figure 10 shows the calculated pressure dependence of the equivalent pseudo-cubic lattice parameters for the orthorhombic phase of (a) $CaMnO_3$ and (b) $PrMnO_3$ using GGA and comparison with experimental data reported in the literature [13, 54-55]. The computed bulk modulus for $CaMnO_3$ and $PrMnO_3$ are found to be 185.4 (Expt. 224 ± 25) [54] GPa and 137.4 (Expt. 139±4) [13] GPa, respectively. It is evident from the figure that compression is more anisotropic for $PrMnO_3$ as compared to $CaMnO_3$. The pressure dependence of the Jahn-Teller distortion, which is evidenced by the spatial distribution of Mn-O bond length (d) and defined as $\Delta= (\frac{1}{6})\sum_{n=1}^{6}[\frac{d_n-<d>}{<d>}]^2$), along-with three Mn-O distances of the distorted MnO$_6$ octahedra, are shown in Figs. 11 (c) and (d). It is interesting to notice that computed Jahn-Teller distortion using the above relation was found to be two-orders lower for CaMnO$_3$ in comparison to PrMnO$_3$, and could be considered as zero. This result is satisfying as CaMnO$_3$ is not known to possess [14] Jahn-Teller distortion.

The calculated variations of Mn-O distances are isotropic for CaMnO$_3$ and found to be highly anisotropic for PrMnO$_3$. On increasing pressure, long Mn-O2 (l) distance decrease faster than the other (Mn-O2(s) and Mn-O1) and becomes nearly equal at around 45 GPa. This nearly isotropic behavior of the Mn-O distance at high pressure is attributed to disappearance of Jahn-Teller distortion (similar to CaMnO$_3$). Basically, application of external pressure opens the Mn-O-Mn angles and in turn shortens the Mn-O bond lengths, leading to less distorted octahedra. The calculated Mn magnetic moment in the equilibrium structure at Mn site is 2.6 µ$_B$, which is in agreement with the experimental values of free Mn$^{3+}$ ( 3.0). We found that at about 45 GPa the Mn magnetic moment decrease to 1.0 µ$_B$ which could be signature for insulator to metal transition.

Recently, Mota *et al.*[13] reported a high pressure study of orthorhombic rare-earth manganites using a combination of synchrotron x ray diffraction and Raman scattering technique. The authors observed change in the diffraction patterns of PrMnO$_3$ and disappearance of the Raman spectrum with pressure. The authors [13] suspected that the changes might be related to the structural phase transition from orthorhombic to tetragonal (*I4/mcm*) structure at 45 GPa. However the disappearance of Raman peak could also be due to metal to insulator transition. Hence the first principle simulation with pressure will help to overcome this ambiguity. Theoretically, we can predict the transition pressure by comparing the enthalpy as a function of pressure for different phases. Figure 12 (a) depicts the computed pressure dependence of the difference in the enthalpy ΔH= H$_{ortho}$-H$_{tetra}$ of orthorhombic and tetragonal phases. We find that at low pressure (Pc=19 GPa) orthorhombic phase has the lowest enthalpy (ΔH), as it is the well-known ground state of PrMnO$_3$ at T=0 K. However at pressure above 19 GPa, the tetragonal phase becomes favorable over the orthorhombic phase. The value of the pressure is in agreement to pressure where shear strain becomes unstable (see Fig. 7 of ref.13). Mota *el al.* [13] reported that the strain analysis does not give any evidence for the suppression of the Jahn-Teller distortion at this pressure. If we recall Fig. 11 (d), we find that computed Jahn-Teller distortion decrease rapidly up to this pressure and then later it decreases slowly. Based on combined diffraction and Raman data, Mota *el al.* [13] reported that PrMnO$_3$ undergoes orthorhombic to tetragonal structural phase transition at around 45 GPa. As mentioned above the calculated pressure dependence of the enthalpy suggests that tetragonal phase is energetically favorable over orthorhombic phase above P= 19 GPa. In order to investigate the dynamical stability of tetragonal phase, we computed the phonon dispersion relation (Fig. 12 (b)) in the entire Brillouin zone at 30 GPa. The presence of unstable phonon modes in the tetragonal phase, clearly suggests that although tetragonal phase is energetically favor over orthorhombic phase, it is dynamically unstable. This rules out the possibility of orthorhombic to tetragonal phase transition in $PrMnO_3$ around 45 GPa.

As shown in Fig. 11 the suppression of Jahn-Teller distortion occurs at 45 GPa. Further we found that the magnetic moment at Mn site is also quenched at the same pressure. The suppression of the Jahn-Teller distortion is expected to decrease the volume at high pressures. Thus change in the volume of the unit cell in the experimental data at around 45 GPa may be associated with disappearance of Jahn-Teller distortion followed by insulator to metal transition. Our band structure calculation also suggests that $PrMO_3$ becomes poor metal above 45 GPa. The similar observation is also found in $LaMnO_3$ [56], which show the suppression of Jahn-Teller distortion and insulator to metal transition by application of pressure. In $LaMnO_3$, Jahn-Teller distortion and orbital ordering are known to be completely suppressed well below the insulator to metal transition. In contrast to $LaMnO_3$, in the present case, we notice that suppression of Jahn-Teller distortion and insulator to metal transition occurs simultaneously.

## V. CONCLUSIONS

We have reported inelastic neutron scattering measurements of the $CaMnO_3$ and $PrMnO_3$ in a wide temperature range up to 1251 K. The excitations at 20 meV and 17 meV are found to be associated with the magnetic origin for $CaMnO_3$ and $PrMnO_3$ respectively. The neutron inelastic spectra also show changes across the magnetic as well as structural phase transitions temperatures in both the compounds. Measurements show that in spite of similar structure the presence of Jahn-Teller distortion $PrMnO_3$ might results in broad peaks in the phonon spectra in comparison to $CaMnO_3$ where it exhibits well-defined peaks. The ab-initio phonon calculations are used for analysis of the experimental neutron inelastic spectra. The unstable phonon modes in the cubic *(Pm-3m)* phase may lead to a stabilization of the tetragonal *(I4/mcm)* phase in $AMnO_3$ (A= Ca and Pr).

We also discussed the effect of pressure on the structural distortions in orthorhombic phase of $CaMnO_3$ and $PrMnO_3$. On application of pressure, the variations of Mn-O distances are isotropic for $CaMnO_3$ and highly anisotropic for $PrMnO_3$. Theoretical calculation for $PrMnO_3$ suggests that suppression of Jahn-Teller distortion and insulator to metal transition occurs simultaneously, which is in-contrast to $LaMnO_3$. We found that at high pressure above 20 GPa tetragonal phase in $PrMnO_3$ is energetically favored over orthorhombic phase, however it is found to be dynamically unstable.

**Acknowledgements**

The neutron scattering experiments at Oak Ridge National Laboratory's Spallation Neutron Source were sponsored by the Scientific User Facilities Division, Office of Basic Energy Sciences, U.S. Department of Energy.

TABLE I. Comparison of experimental and theoretical structural parameters in the antiferromagnetic orthorhombic phase (space group: *Pnma*) of CaMnO$_3$. The Wyckoff sites of the atoms are given in the brackets. "FRM", "FRNM" and "PNM" refer to fully relaxed magnetic, fully relaxed non-magnetic and partially relaxed non magnetic calculations, respectively.

| Structural Parameters | Expt. [18] | Ref. [33] | **This work** | | | | | |
|---|---|---|---|---|---|---|---|---|
| | | | GGA | | | LDA | | |
| | | | FRM | PNM | FRNM | FRM | PNM | FRNM |
| $A_o$ (Å) | 5.279 | 5.287 | 5.3380 | 5.3380 | 5.3148 | 5.2011 | 5.2011 | 5.189 |
| $B_o$ (Å) | 7.448 | 7.498 | 7.4977 | 7.4977 | 7.4238 | 7.2933 | 7.2933 | 7.2386 |
| $C_o$ (Å) | 5.264 | 5.235 | 5.2949 | 5.2949 | 5.2645 | 5.1396 | 5.1396 | 5.1084 |
| Ca (4c) | | | | | | | | |
| x | 0.035 | 0.040 | 0.040 | 0.038 | 0.038 | 0.045 | 0.044 | 0.045 |
| y | 0.250 | 0.250 | 0.250 | 0.250 | 0.250 | 0.250 | 0.250 | 0.250 |
| z | -0.009 | -0.008 | -0.008 | -0.007 | -0.007 | -0.009 | -0.008 | -0.009 |
| Mn (4b) | | | | | | | | |
| x | 0.00 | 0.00 | 0.00 | 0.00 | 0.00 | 0.00 | 0.00 | 0.00 |
| y | 0.00 | 0.00 | 0.00 | 0.00 | 0.00 | 0.00 | 0.00 | 0.00 |
| z | 0.500 | 0.500 | 0.500 | 0.500 | 0.500 | 0.500 | 0.500 | 0.500 |
| O1 (4c) | | | | | | | | |
| x | 0.493 | 0.485 | 0.489 | 0.483 | 0.483 | 0.488 | 0.491 | 0.491 |
| y | 0.250 | 0.250 | 0.250 | 0.250 | 0.250 | 0.250 | 0.250 | 0.250 |
| z | 0.068 | 0.071 | 0.068 | 0.062 | 0.062 | 0.073 | 0.070 | 0.070 |
| O2 (8d) | | | | | | | | |
| x | 0.290 | 0.287 | 0.288 | 0.287 | 0.287 | 0.290 | 0.290 | 0.29 |
| y | 0.030 | 0.036 | 0.035 | 0.032 | 0.033 | 0.031 | 0.036 | 0.037 |
| z | -0.289 | -0.288 | -0.289 | -0.288 | -0.288 | -0.289 | -0.289 | -0.289 |

TABLE II. The calculated structures of CaMnO$_3$ in the tetragonal (*I4/mcm*) and super cell ($\sqrt{2}\times\sqrt{2}\times2$) of cubic phase. The super cell ($\sqrt{2}\times\sqrt{2}\times2$) of cubic phase is equivalent to the tetragonal (I4/mcm) phase. The unit cell in the *I4/mcm* space group has O1, O2, Ca and Mn atoms at the 4a(0 0 0.25), 8h (x y 0), 4b (0.5 0 0.25) and 4c(0 0 0) Wyckoff sites respectively. The distortion vector is obtained from the difference in atomic coordinates of the tetragonal phase (I4/mcm) and super cell cubic phase. The eigen vector of the unstable R point mode in the super cell ($\sqrt{2}\times\sqrt{2}\times2$) of cubic phase (Pm-3m) is also given. The amplitude of the eigen vector of O2 is scaled to match with the distortion vector.

| Atom |   | Super cell ($\sqrt{2}\times\sqrt{2}\times2$) of cubic phase (*Pm-3m*) | Tetragonal phase (*I4/mcm*) | Distortion vector in fractional coordinates | Eigen vector of the unstable R point mode in the super cell of cubic phase |
|---|---|---|---|---|---|
|  | $a_t$ (Å) | 5.315 (Å) | 5.279 |  |  |
|  | $c_t$ (Å) | 7.518 (Å) | 7.621 |  |  |
| O1 | x | 0.00 | 0.00 | 0.00 | 0.00 |
|  | y | 0.00 | 0.00 | 0.00 | 0.00 |
|  | z | 0.25 | 0.25 | 0.00 | 0.00 |
| O2 | x | 0.25 | 0.31 | 0.06 | 0.06 |
|  | y | 0.25 | 0.19 | -0.06 | -0.06 |
|  | z | 0.00 | 0.00 | 0.00 | 0.00 |
| Ca | x | 0.50 | 0.50 | 0.00 | 0.00 |
|  | y | 0.00 | 0.00 | 0.00 | 0.00 |
|  | z | 0.25 | 0.25 | 0.00 | 0.00 |
| Mn | x | 0.00 | 0.00 | 0.00 | 0.00 |
|  | y | 0.00 | 0.00 | 0.00 | 0.00 |
|  | z | 0.00 | 0.00 | 0.00 | 0.00 |

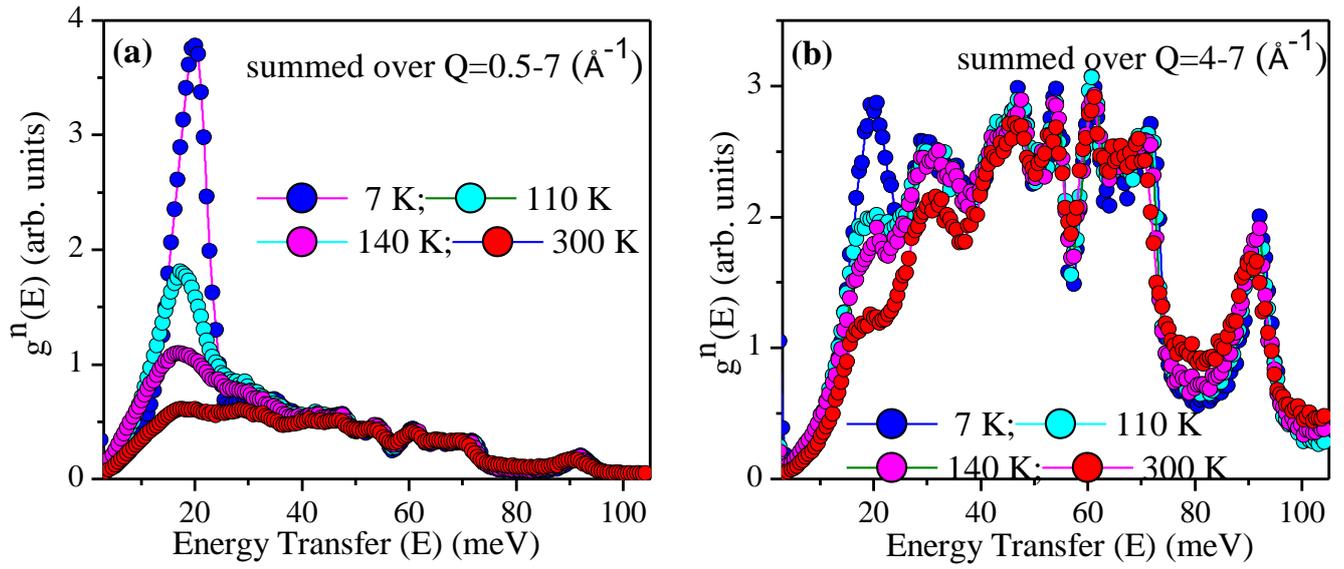

**FIG 1.** (Color online) The neutron inelastic spectra of $CaMnO_3$ at low temperatures, the data were summed over (a) Q=0.5-7 Å$^{-1}$ and (b) Q=4-7 Å$^{-1}$ respectively. The peak at ~20 meV is due to spin-wave excitations, not due to phonons (1 meV=8.0585 cm$^{-1}$).

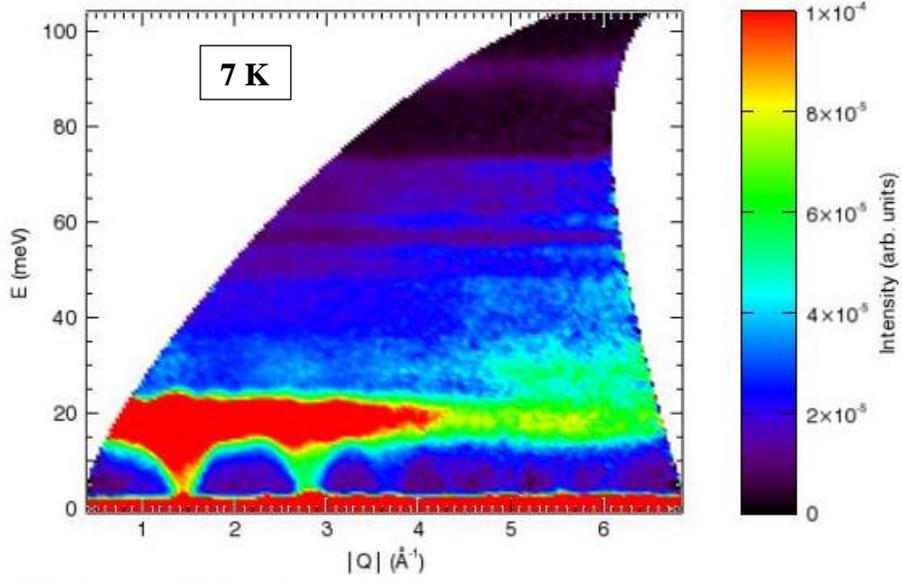
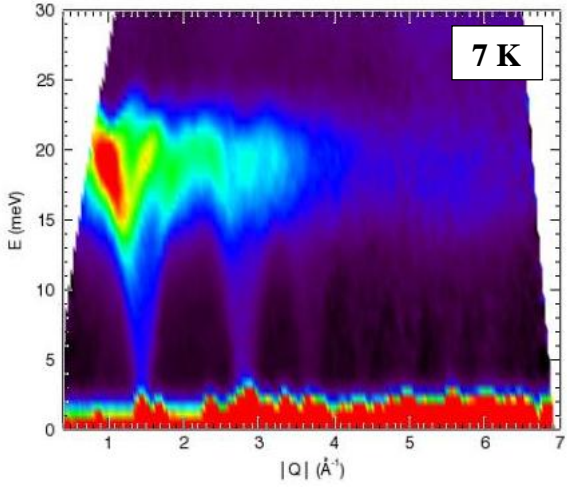
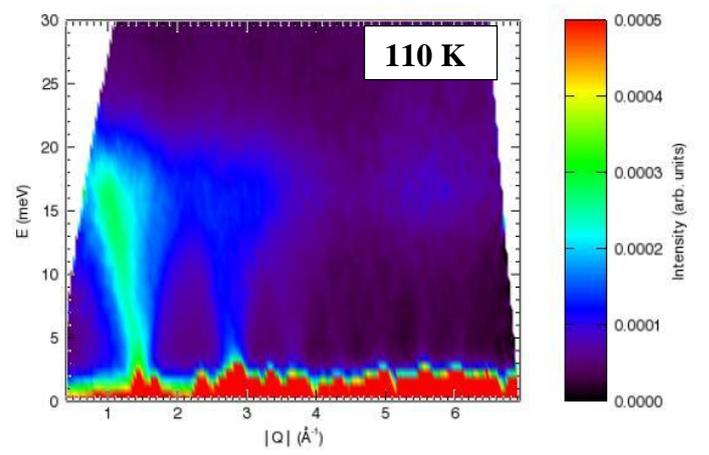
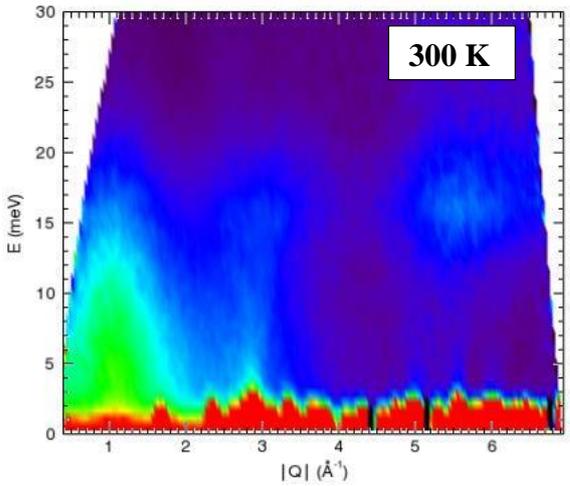
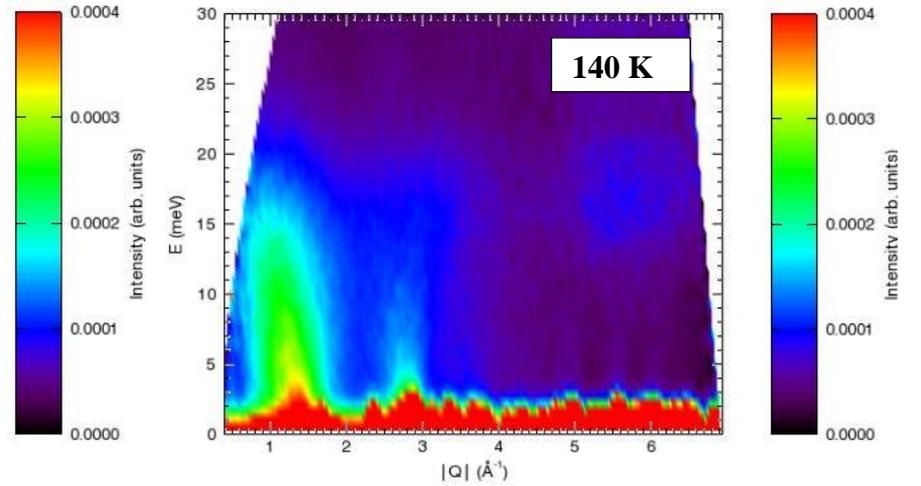

FIG 2. (Color online) The (Q,E) contour plot of S(Q,E) data for CaMnO$_3$ at T=7 K measured at SEQUOIA with incident neutron energy of 110 meV is shown at top. Strong intensity excitations at low temperatures (7 K and 110 K) below E=20 meV and Q=3.5 Å$^{-1}$ are due to magnetic spin-wave excitations. The excitations around 30, 45, 55, 60, 65, 70, and 90 meV are due to phonons (their intensities increase with increasing Q).

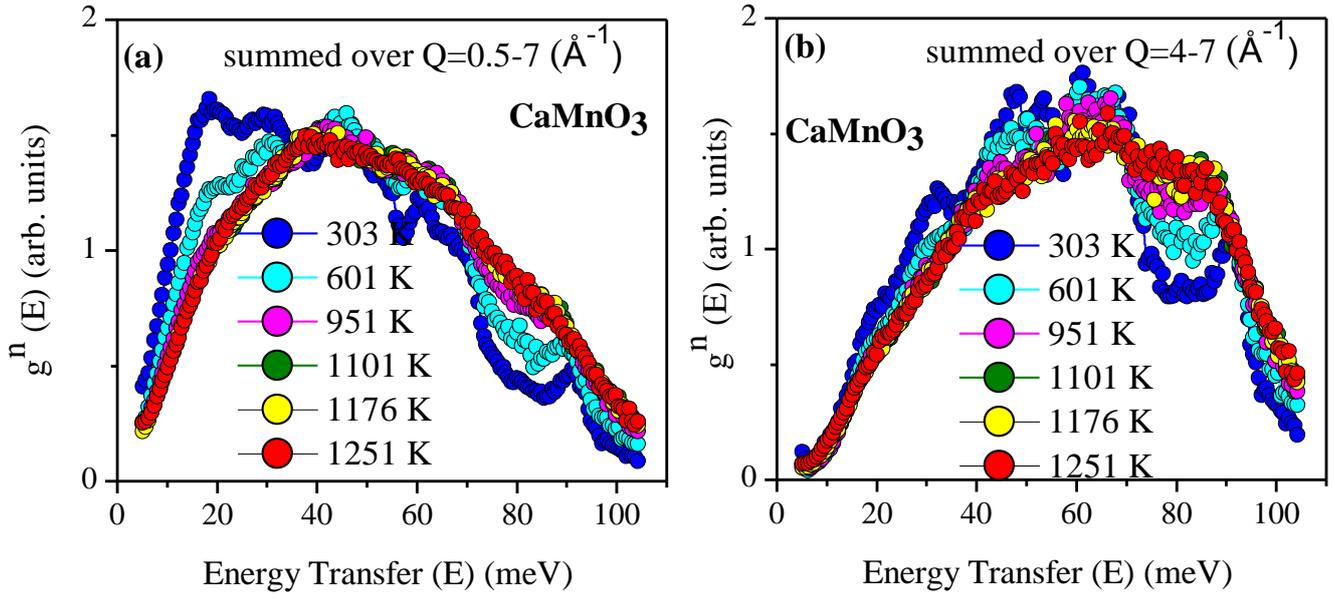

FIG 3. (Color online) The temperature dependence (above 300 K) of the neutron inelastic spectra of CaMnO$_3$, the data were summed over (a) Q=0.5-7 Å$^{-1}$ and (b) Q=4-7 Å$^{-1}$ respectively.

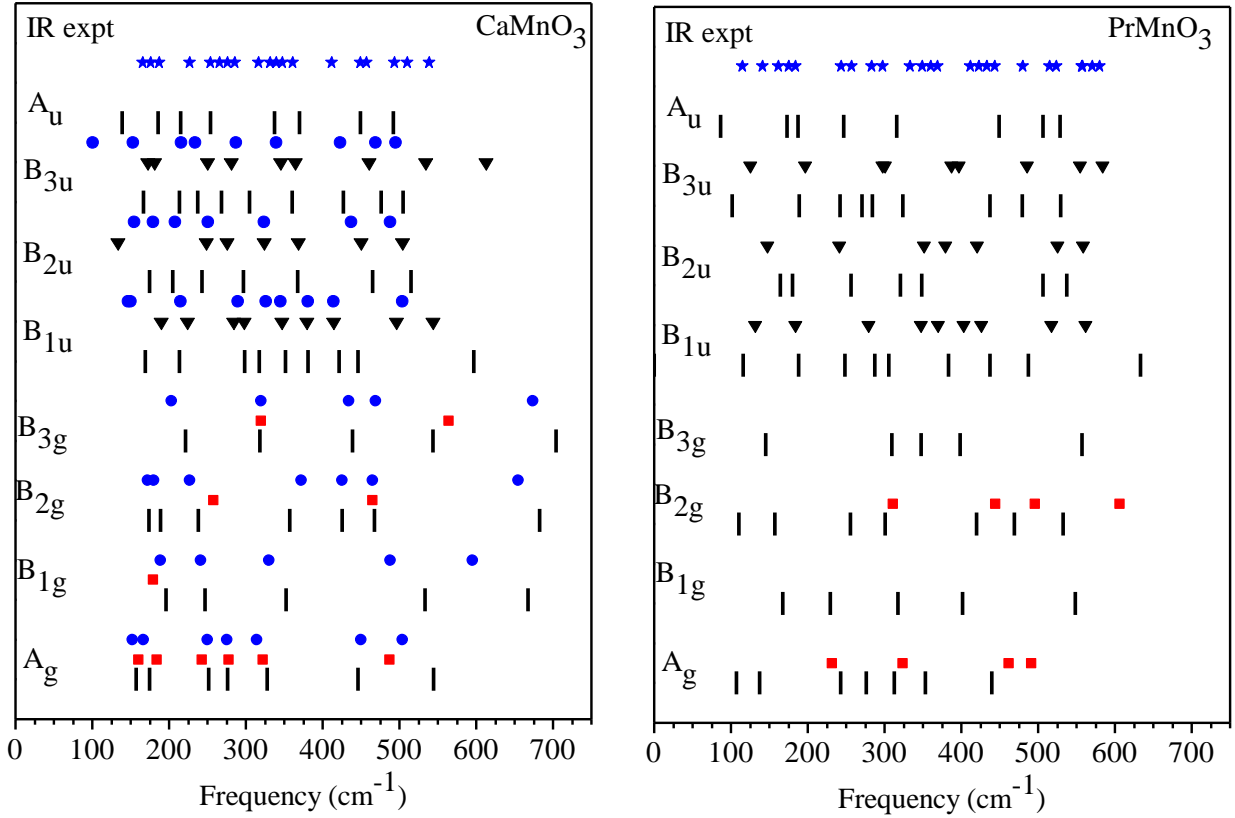

FIG 4. (Color online) Comparison of the calculated long-wavelength phonon frequencies with the available data in literature. The references for reported experimental and calculated optical long-wavelength data are as [a] Raman experimental data (ref. 51 for $CaMnO_3$; ref. 32 for $PrMnO_3$), [b] first principle DFT calculation (ref. 33 for $CaMnO_3$), [c] IR experimental data (ref. 52 for $CaMnO_3$ and $PrMnO_3$) and (d) lattice dynamical calculation using potential model (ref. 51 for $CaMnO_3$; ref. 32 for $PrMnO_3$).

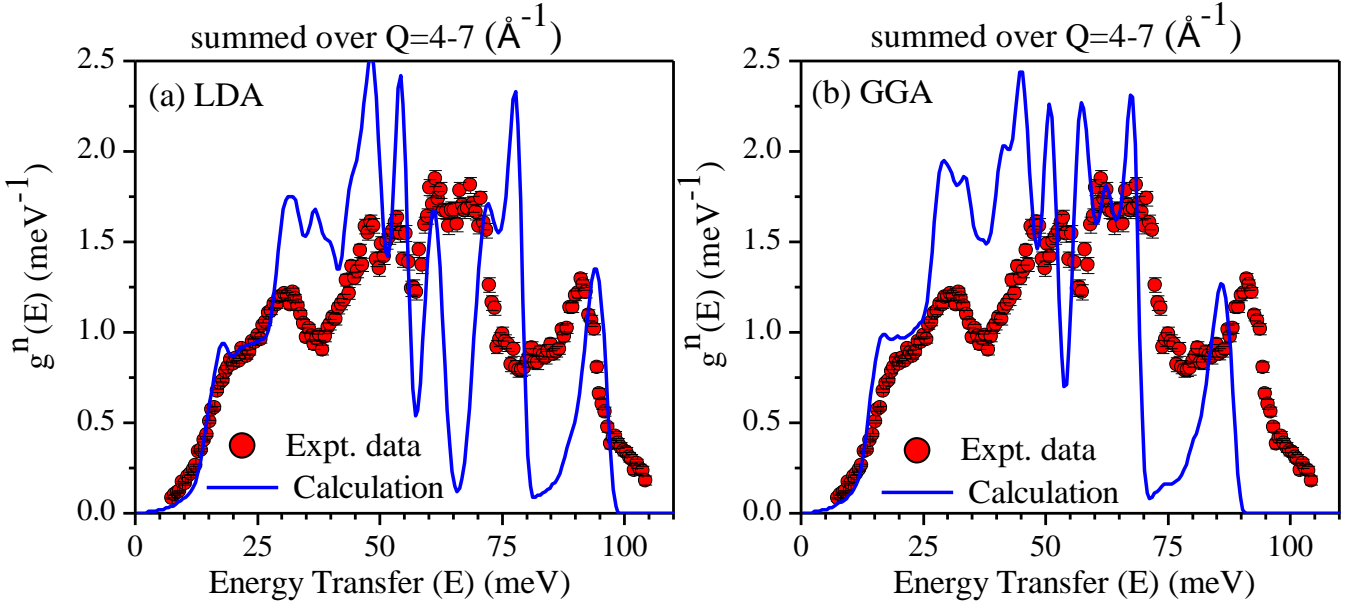

FIG 5. (Color online) Comparison between the experimental (T=300 K) and calculated neutron inelastic spectra of CaMnO$_3$ using (a) local density approximation and (b) generalized gradient approximation. Experimental data are summed over 4-7 Å$^{-1}$. The phonon calculations are carried out in the fully relaxed magnetic (FM) configuration. The calculated phonon spectra have been convoluted with a Gaussian of FWHM of 4.5 meV to account for the effect of energy resolution in the experiment.

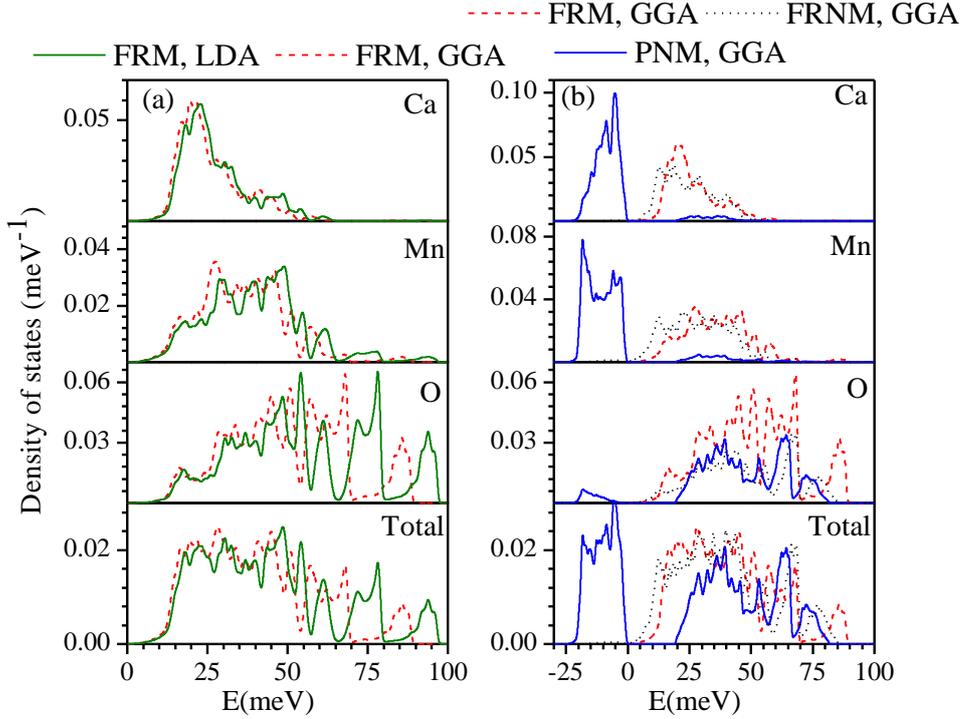

FIG 6. (Color online) (a) The calculated partial phonon density of states of various atoms in CaMnO$_3$ with in LDA and GGA approximations. (b) The calculated partial density of states of CaMnO$_3$ in various configurations with in GGA. "FRM", "FNM" and "PNM" refer to fully relaxed magnetic, fully relaxed non-magnetic and partially relaxed non magnetic calculations, respectively. The energies of unstable modes in PNM-GGA are plotted as negative energies.

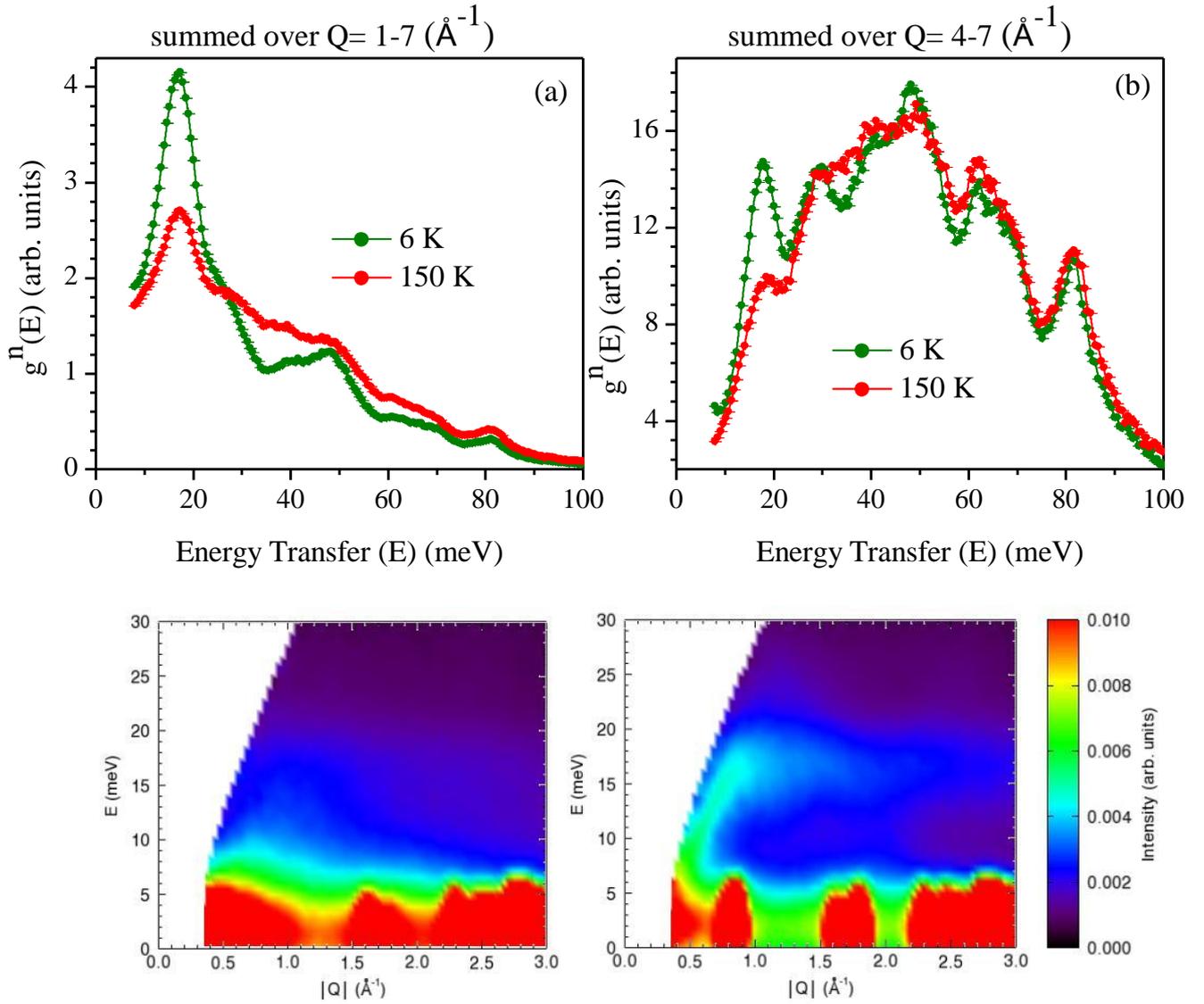

FIG 7. (Color online) Top panel: Temperature dependent neutron inelastic spectra of PrMnO$_3$ summed over various Q-range. Bottom panel: Contour plot of S(Q,E) spectra for PrMnO$_3$ measured at 6 K (right) and 150 K (left). A dispersed spin wave excitation is clearly seen below 20 meV and 1.5 Å$^{-1}$ at 6 K. In 150 K spectra, weakly dispersed magnetic excitation around 15 meV is observed.

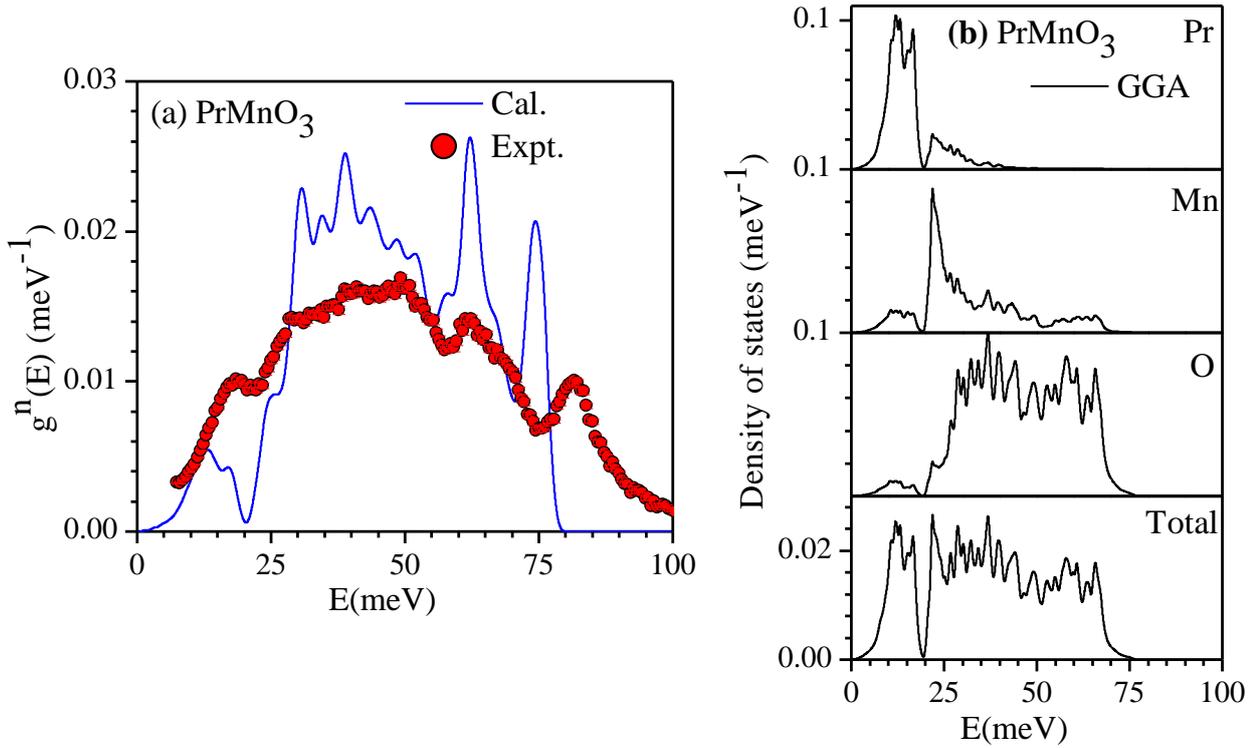

FIG 8. (Color online) (a) Comparison between the experimental (T= 150 K) and calculated phonon spectra in PrMnO$_3$. Experimental data are summed over 4-7 Å$^{-1}$. (b) The calculated partial phonon density of states of various atoms in PrMnO$_3$. The phonon calculations are carried out in the fully relaxed magnetic (FRM) configuration in the generalized gradient approximation (GGA).

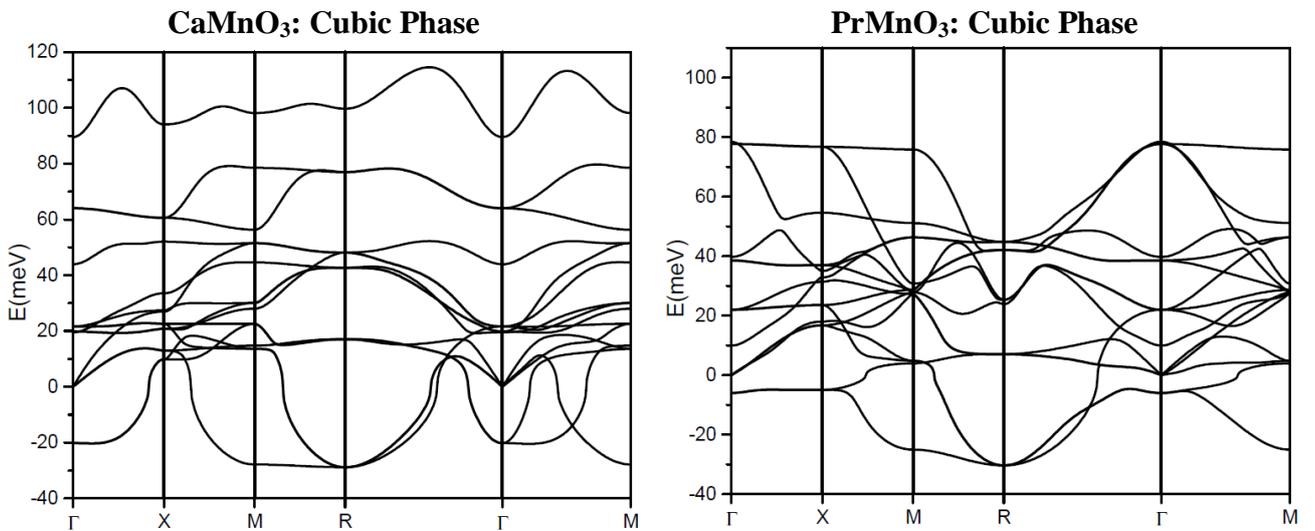

FIG 9. Computed phonon dispersion relations for CaMnO$_3$ (left panel) and PrMnO$_3$ (right panel) using generalized gradient approximation (GGA) in cubic phase.

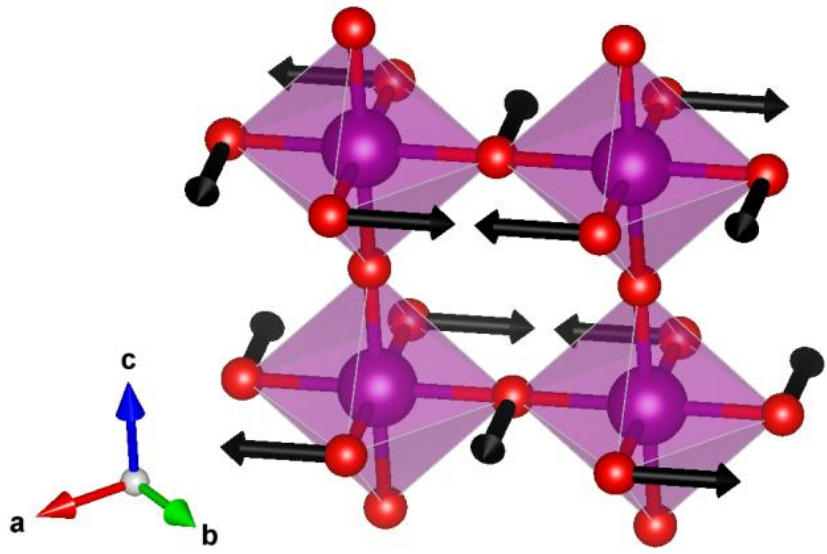

FIG 10. (Color online) The displacements pattern of unstable mode at the R point in the cubic phase.

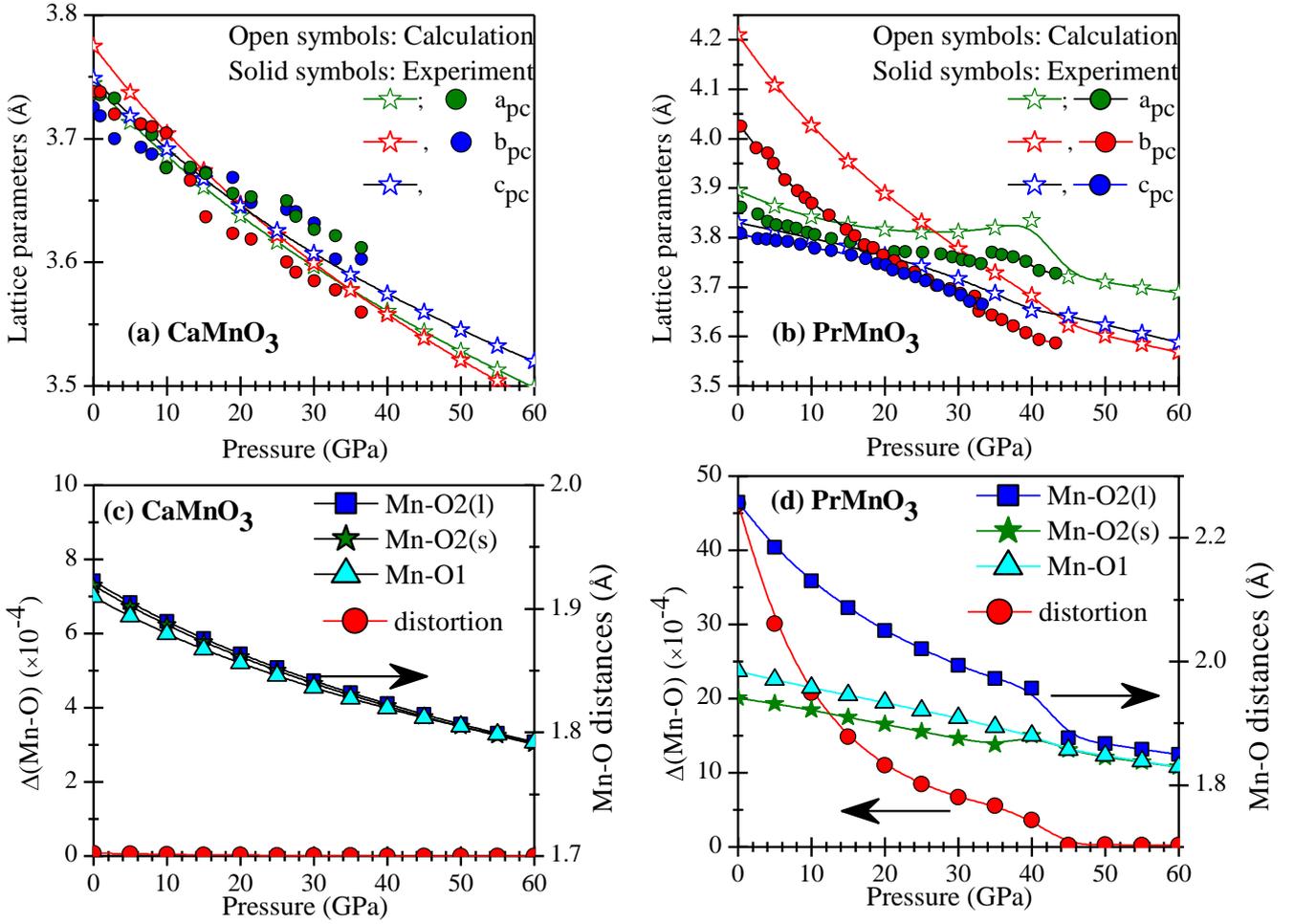

FIG 11. (Color online) Pressure dependence of pseudocubic lattice parameters for (a) CaMnO$_3$ and (b) PrMnO$_3$ compared to reported experimental data for CaMnO$_3$ [54] and PrMnO$_3$ [13] respectively. Pressure dependence of Mn-O bond length and distortion of MnO$_6$ ochtahedra as calculated are shown in (c) CaMnO$_3$ and (d) PrMnO$_3$, respectively.

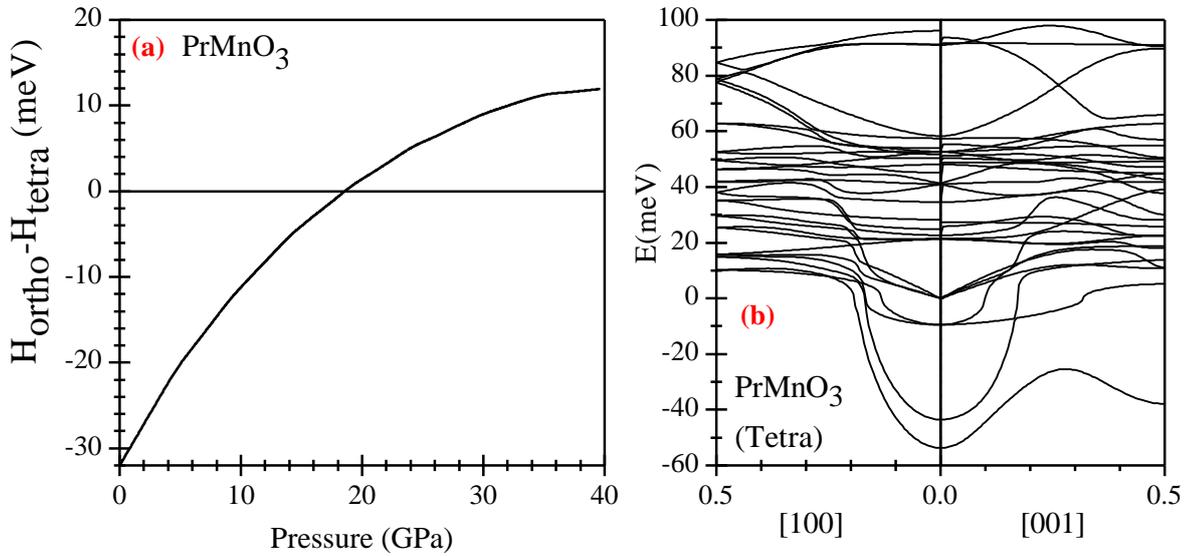

FIG 12. (a: ) Enthalpy difference ($\Delta H$) between the orthorhombic (*Pnma*) and tetragonal (*I4/mcm*) phases of PrMnO$_3$ as calculated using ab-inito DFT calculation. The computed phonon dispersion relations for PrMnO$_3$ in tetragonal phase at P= 30 GPa is shown in (b).